\documentclass[12pt]{article}
\usepackage[dvips]{graphicx}
\usepackage[tight]{subfigure}
\usepackage{amssymb}

 \usepackage[utf8]{inputenc}
\usepackage{amsmath}
\usepackage{amsfonts}
\usepackage{amssymb}
\usepackage{authblk}

\usepackage{bm}
\usepackage{empheq}
\usepackage{epsfig}
\usepackage{graphicx}
\usepackage{epstopdf}
\usepackage{mathtools}
\usepackage{amssymb}
\usepackage{amsmath}

\makeindex
\def\disp{\displaystyle}

\def\be{\begin{equation}}
\def\ee{\end{equation}}
\def\bea{\begin{eqnarray}}
\def\eea{\end{eqnarray}}
\def\beaN{\begin{eqnarray*}}
\def\eeaN{\end{eqnarray*}}
\def\ed{\end{document}}
\def\bit{\begin{itemize}}
\def\eit{\end{itemize}}

\def\sig{\sigma}
\def\Sig{\Sigma}
\def\lam{\lambda}
\def\Lam{\Lambda}
\def\Del{\Delta}
\def\del{\delta}

\def\k{\kappa}

\def\alf{\alpha}

\def\BN{\bar\nabla}

\def\3nabla{{\stackrel{\scriptscriptstyle 3}{\nabla}}}
\def\2nabla{{\stackrel{\scriptscriptstyle 2}{\nabla}}}

\def\di{\partial}

\def\bn{\bar\nabla}

\def\half{{\textstyle{1 \over 2}}}

\def\~{\tilde}
\def\lag{{{\cal L}}}

\def\m{\label}
\def\l{\left}
\def\r{\right}

\def\goto{\rightarrow}

\def\const{\rm const}
\def\diag{\rm diag}

\def\citep{\cite}

\usepackage{cite}

\newcommand{\nc}{\newcommand}

\nc{\bse}{\begin{equation*}}
\nc{\ese}{\end{equation*}}
\nc{\ba}{\begin{array}}
\nc{\ea}{\end{array}}
\nc{\bal}{\begin{align}}
\nc{\eal}{\end{align}}
\nc{\bi}{\begin{description}}
\nc{\ei}{\end{description}}
\nc{\Def}{=}
\nc{\sign}{\mathrm{sign}\;}
\nc{\diagR}{\mathrm{diag}\;}
\nc{\constR}{\mathrm{const}}
\nc{\Tr}{\mathrm{Tr}\,}
\nc{\id}{\mathrm{id}}
\nc{\eq}{\equiv}
\nc{\we}{\wedge}
\nc{\ra}{\rightarrow}
\nc{\bfrac}{\disp\frac}

\nc{\bfpa}{{\bm{\partial}}}                       
\nc{\pa}[1]{{\partial_{#1}}{}}                    
\nc{\pau}[1]{{\partial^{#1}}{}}                   
\nc{\alp}{\alpha}
\nc{\bet}{\beta}
\nc{\gam}{\gamma}
\nc{\eps}{\epsilon}
\nc{\veps}{\varepsilon}
\nc{\zet}{\zeta}
\nc{\tet}{\theta}
\nc{\vtet}{\vartheta}
\nc{\iot}{\iota}
\nc{\kap}{\kappa}
\nc{\vkap}{\varkappa}
\nc{\vpi}{\varpi}
\nc{\vrho}{\varrho}
\nc{\vsig}{\varsigma}
\nc{\ups}{\upsilon}
\nc{\vphi}{\varphi}
\nc{\ome}{\omega}

\nc{\Gam}{\Gamma}
\nc{\Tet}{\Theta}
\nc{\Ups}{\Upsilon}
\nc{\Ome}{\Omega}

\nc{\BL}[1]{\bm{#1}}

\nc{\bfa}{\BL{a}}
\nc{\bfb}{\BL{b}}
\nc{\bfc}{\BL{c}}
\nc{\bfd}{\BL{d}}
\nc{\bfe}{\BL{e}}
\nc{\bff}{\BL{f}}
\nc{\bfg}{\BL{g}}
\nc{\bfh}{\BL{h}}
\nc{\bfi}{\BL{i}}
\nc{\bfj}{\BL{j}}
\nc{\bfk}{\BL{k}}
\nc{\bfl}{\BL{l}}
\nc{\bfm}{\BL{m}}
\nc{\bfn}{\BL{n}}
\nc{\bfo}{\BL{o}}
\nc{\bfp}{\BL{p}}
\nc{\bfq}{\BL{q}}
\nc{\bfr}{\BL{r}}
\nc{\bfs}{\BL{s}}
\nc{\bft}{\BL{t}}
\nc{\bfu}{\BL{u}}
\nc{\bfv}{\BL{v}}
\nc{\bfw}{\BL{w}}
\nc{\bfx}{\BL{x}}
\nc{\bfy}{\BL{y}}
\nc{\bfz}{\BL{z}}

\nc{\bfA}{\BL{A}}
\nc{\bfB}{\BL{B}}
\nc{\bfC}{\BL{C}}
\nc{\bfD}{\BL{D}}
\nc{\bfE}{\BL{E}}
\nc{\bfF}{\BL{F}}
\nc{\bfG}{\BL{G}}
\nc{\bfH}{\BL{H}}
\nc{\bfI}{\BL{I}}
\nc{\bfJ}{\BL{J}}
\nc{\bfK}{\BL{K}}
\nc{\bfL}{\BL{L}}
\nc{\bfM}{\BL{M}}
\nc{\bfN}{\BL{N}}
\nc{\bfO}{\BL{O}}
\nc{\bfP}{\BL{P}}
\nc{\bfQ}{\BL{Q}}
\nc{\bfR}{\BL{R}}
\nc{\bfS}{\BL{S}}
\nc{\bfT}{\BL{T}}
\nc{\bfU}{\BL{U}}
\nc{\bfV}{\BL{V}}
\nc{\bfW}{\BL{W}}
\nc{\bfX}{\BL{X}}
\nc{\bfY}{\BL{Y}}
\nc{\bfZ}{\BL{Z}}

\nc{\bfalp}{\bm{\alp}}
\nc{\bfbet}{\bm{\bet}}
\nc{\bfgam}{\bm{\gam}}
\nc{\bfdel}{\bm{\del}}
\nc{\bfeps}{\bm{\eps}}
\nc{\bfveps}{\bm{\veps}}
\nc{\bfzet}{{\bm{\zet}}}
\nc{\bfeta}{\bm{\eta}}
\nc{\bftet}{\bm{\tet}}
\nc{\bfvtet}{\bm{\vtet}}
\nc{\bfiot}{\bm{\iot}}
\nc{\bfkap}{\bm{\kap}}
\nc{\bfvkap}{\bm{\vkap}}
\nc{\bflam}{\bm{\lam}}
\nc{\bfmu}{\bm{\mu}}
\nc{\bfnu}{\bm{\nu}}
\nc{\bfxi}{\bm{\xi}}
\nc{\bfpi}{\bm{\pi}}
\nc{\bfvpi}{\bm{\vpi}}
\nc{\bfrho}{\bm{\rho}}
\nc{\bfvrho}{\bm{\vrho}}
\nc{\bfsig}{\bm{\sig}}
\nc{\bfvsig}{\bm{\vsig}}
\nc{\bftau}{\bm{\tau}}
\nc{\bfups}{\bm{\ups}}
\nc{\bfphi}{\bm{\phi}}
\nc{\bfvphi}{\bm{\vphi}}
\nc{\bfchi}{\bm{\chi}}
\nc{\bfpsi}{\bm{\psi}}
\nc{\bfome}{\bm{\ome}}

\nc{\bfGam}{\bm{\Gam}}
\nc{\bfDel}{\bm{\Del}}
\nc{\bfTet}{\bm{\Tet}}
\nc{\bfLam}{\bm{\Lam}}
\nc{\bfXi}{\bm{\Xi}}
\nc{\bfPi}{\bm{\Pi}}
\nc{\bfSig}{\bm{\Sig}}
\nc{\bfUps}{\bm{\Ups}}
\nc{\bfPhi}{\bm{\Phi}}
\nc{\bfPsi}{\bm{\Psi}}
\nc{\bfOme}{\bm{\Ome}}

\begin{document}

\title{\bf Black holes of the Vaidya type with flat and (A)dS asymptotics as point particles}

\author{A. N. Petrov\thanks{Electronic address: \texttt{alex.petrov55@gmail.com}}}

\affil{Sternberg Astronomical institute, MV Lomonosov Moscow state university  \protect\\ Universitetskii pr. 13, Moscow, 119992, Russia}

\date{\small \today}
\maketitle


\begin{abstract}
A presentation of the Vaidya type Schwarzschild-like black holes with flat, AdS and dS asymptotics in 4-dimensional general relativity in the form of a pointlike mass is given. True singularities are described by making the use of the Dirac $\delta$-function in a non-contradictory way. The results essentially generalize previous derivations where the usual Schwarzschild black hole solution is represented in the form of a point particle. The field-theoretical formulation of general relativity, which is equivalent to its standard geometrical formulation, is applied as an alternative mathematical formalism. Then perturbations on a given background are considered as dynamical fields propagating in a given (fixed) spacetime. The energy (mass) distribution of such field configurations is just represented as a point mass. The new description of black holes' structure can be useful in explaining and understanding their features and can be applied in  calculations with black hole models. A possibility of application of the field-theoretical formalism in studying the regular black hole solutions is discussed.
\end{abstract}

\medskip


\medskip

\section{Introduction}
\m{Introduction}

 Many solutions in general relativity (GR) and other metric theories include singularities in curvature. Among them there are black hole solutions like the Schwarzschild, Kerr, Kerr-Newman etc., one has to note also solutions with cosmic strings, and many others. Usually, basing on the standard formalism, authors have a convention that the field gravitational equations do not hold at the singularities. In spite of that it is clear that such a convention is not so fruitful. Therefore many authors give a description by the way when singularities are presented with making the use of generalized functions, like Dirac's $\delta$-function, $\theta$-function, Green's function. They could be incorporated into the energy-momentum of sources, which could be interpreted as a matter source of a curved geometry with defects. Singularities can be taken into account in a special representation of the curvature distribution as well, besides authors use regularization methods. In all the cases after such procedures the interpretation has being developed in the way when the field equations hold in whole spacetime including singularities. There is a wide literature related to this topic, we would suggest to see the papers \cite{ADM_1960,ADM_1960a,Tangherlini_1961,Parker_1979,Taub_1980,Raju_1982,Colombeau_1983,Geroch_Traschen_1987,Balasin_Nachbagauer_1994,Clarke_Vickers_Wilson_1996,Wilson_1997,Vickers_Wilson_1999,Pantoja_Rago_2002,Melis_Mignemi_2005,Cadoni_Mignemi_2005,Lundgren_Schmekel_York_2007} and references therein. {Nowadays, such constructions are continued. Recently, in \cite{Sazhina_2023} a curved cosmic string with making the use of Dirac's $\delta$-function is suggested. It is a more complicated case with respect a straight cosmic string considered in \cite{Sokolov_Starobinsky_1977}.}

 The Schwarzschild solution \cite{Misner_Thorn_Wheeler_1973} takes the modest place in the list of solutions with singularities. However, in spite of its simplicity, it is quite rich and interesting.
 The Schwarzschild solution has been obtained under the simplest assumptions: it is a vacuum spherically symmetric static solution. The mass parameter $M$ has an interpretation in a  correspondence with the assumption that the solution is induced by isolated masses and has the Newtonian asymptotic behaviour. To resolve Einstein equations under such assumptions one chooses a matter source as a point particle (point mass). But, in the result, one finds that the Schwarzschild solution presents a black hole solution with a complicated geometry including the event horizon and the spacelike true singularity that is not a point particle. Thus, the notion of a point mass in the form of the source in GR (and many other metric theories) becomes unclear. Many researches studied this problem trying to preserve a notion of a point particle, see, for example, \cite{Bel_1969,Kawai_Sakane_1997,Heinzle_2002,Katanaev_2013,Fiziev_2019} and references therein. Particularly, a short review of the aforementioned papers can be found in our works \cite{Petrov_Narlikar_1996,Petrov_2005,Petrov_2018}.

To find out a pointlike description for the Schwarzschild solution it is useful to keep in mind how a point mass is described in the Newtonian gravity. Recall that the equation of the Newtonian gravity is the Laplace equation. It is working even if the mass density  (mass distribution) is presented by Dirac's $\delta$-function: $\varepsilon = M\delta(\bm r)$. Indeed, the Newtonian equation is satisfied in the whole space including even the point $r = 0$. The volume integration of $\varepsilon$ over the whole space gives the accepted result for the total mass $M$. It is important to remark that the surface integration gives the same result $M$.

Our results in representation of the Schwarzschild black hole \cite{Petrov_Narlikar_1996,Petrov_2005,Petrov_2018} in the form of a massive point particle and with the derivation of the true singularity by Dirac's $\delta$-function are based on the field-theoretical approach, see \cite{GPP,[15],GP87,CQG_DT,Petrov+_2017,Petrov_Pitts_2019}, and references therein. Such a reformulation of GR, which is totally equivalent to its standard geometrical formulation, is applied as a mathematical instrument only. Metric and matter perturbations, which are finite ones (not infinitesimal or approximate), are considered on a given background as dynamical fields (the field configuration). The energy density of perturbations allows us to interpret the Schwarzschild black hole as a pointlike mass distribution. Its form is totally close to the Newtonian derivation of the point source,  see \cite{Petrov_2005,Petrov_2018} and the review \cite{Petrov_Pitts_2019}.

To the best of our knowledge, black holes with de Sitter (dS) or anti-de Sitter (AdS) asymptotics  have been considered without a deep study in the context of representation in the form of a point particle. See, for example, \cite{Cadoni_Mignemi_2005} for the Schwarzschild-AdS black holes where the problem is noted only, and see \cite{Lundgren_Schmekel_York_2007} for the Schwarzschild-dS black holes where the energy is considered as concentrated under the event horizon with the use of the quasi-local technique by Brown and York \cite{Brown_York_1993}. In papers \cite{Petrov_2005,Petrov_2018}, we have considered the standard flat asymptotics for the Newtonian-like point representation and for the continuous collapse to the Newtonian-like point. However, up to now we did not considered Schwarzschild-(A)dS black holes as well.

Recently, we have elaborated the field-theoretical formalism for constructing conserved quantities (currents and superpotentials) in the Lovelock gravity \cite{Lovelock} giving maximally general expressions, see \cite{Petrov_2019,Petrov_Pitts_2019}. Later \cite{Petrov_2021}, we have adopted this technique to a so-called {\em pure} Lovelock gravity (see, for example, \cite{Dadhich+_2013a,Kastikainen_2020} and references therein) where Lagrangian is presented only by a zeroth term (a bare cosmological constant) and a single Lovelock polynomial with a related coupling constant. In \cite{Petrov_2021}, the adopted formalism was applied for constructing conserved quantities for both, the static \cite{Cai_Ohta_2006} and the dynamic Vaydia type \cite{Cai+_2008} Schwarzschild-like black holes with AdS, dS and flat asymptotics in the pure Lovelock gravity.

All the results \cite{Petrov_2021} are general and valid for arbitrary dimensions and for arbitrary order of the single Lovelock polynomial. However, our interpretation of the results in \cite{Petrov_2021} has been accented to general dimensions $D>4$ and to the order of the related single polynomial more than 1. However, the case $D=4$ with the order 1, that presents the usual GR with the cosmological constant (and the pure Lovelock gravity as well), has not been discussed. Nevertheless, namely this case is more intriguing in the context of the pointlike presentation.

Basing on this background, we formulate the main goal of the present paper as to describe the Schwarzschild and Schwarzschild-(A)dS black holes of the Vaidya type with the generalized lightlike matter (null liquid) in the form of a pointlike object. Concerning the null liquid, first, although it is not presented in nature, its properties are close to those of a real matter and, thus, are useful in constructing various models. Second, being simple in derivation, the null liquid allows us to construct models in a more economical way.
Our study is based on the results \cite{Petrov_2021} related to 4-dimensional GR. Because we consider spherically symmetric black holes without matter charges (e.g., electric ones) and rotations, we consider only the energy and the energy flux and related characteristics given by currents and energy-momentum of perturbations.  Thus we try, although particularly, to close the aforementioned gap in reconstructing various types of black holes in the form of a point particle.

We stress again that the equations of GR in the field-theoretical  formulation are equivalent to those in the standard formulation. They are two various mathematical formalisms of the same theory. By this, all the physical and geometrical effects of GR are the same in the both formalisms, see \cite{Petrov+_2017}. For example, considering geodesics in the standard formulation, we use the usual geometry properties defined by the physical metric. Considering a motion of free test particles in the field-theoretical formulation, we use the geometry properties defined by the effective metric that is described by the background metric and perturbations. In both the cases, as a result, one has equivalent predictions.

However, concerning singularities the standard formulation classifies them as specialities where the gravitational equations do not hold. This means that mathematical operations are not defined. On the other hand, if a singularity is modeled by a generalized function one can provide calculations in whole. The bright example of using such mathematical possibilities is as follows. Last time a direct detection of gravitational waves (produced in the events of coalescence of binary black holes) by laser interferometers, LIGO and Virgo, becomes an ordinary event. To be convinced in their interpretation next mathematical modeling is provided. Black holes from the first step of iterations are modeled by pointlike particles presented namely by Dirac's $\delta$-function. Recall that mathematical operations with the $\delta$-function are well defined. Next steps of iterations include more complicated generalized special functions, however all of them allow us to provide also well defined calculations. On such calculations one can see, for example, in \cite{Damour_Jaranowski_Schafer_2001,Blanchet_2014}.  In the present work, $\delta$-function just appears in the field-theoretical description, when the gravitational equations hold at the singularity as well.

Keeping the above in mind, we restrict ourselves by the following natural requirements:
\bit

\item[(i)] The true singularity has to be described by the world line $r=0$ in a background spacetime with making the use of Dirac's $\del({\bm r})$-function.

\item[(ii)] The Schwarzschild-like solution has to be presented in the asymptotically flat or (A)dS form with an appropriate fall-off of potentials at spatial infinity consistent with the Newtonian-like behavior.

\item[(iii)] A so-called ``$\eta$-causality'' is required. It is a property, when the physical light cone is placed inside the background light
cone at all the points of the background spacetime.

\eit
The last requirement is necessary to avoid interpretation difficulties under the field-theoretical
presentation of GR. It means that all of the causally connected
events in the physical spacetime must be described by the right causal
structure of the background spacetime. Notions of the $\eta$-causality have been introduced and considered in \cite{Pitts_Schive_2004}.

The paper is organized as follows. In section \ref{ON_FT_approach}, we give and explain notions of the field-theoretical reformulation of GR which are necessary for the present study.

In section \ref{Vaidya_type}, we derive an arbitrary metric of the Vaidya type in the generalized Eddington-Finkelstein coordinates, define related perturbations on a static background with the same type of the metric also. Then, we derive components of a superpotential of the general form, a related gravitational charge and a flux of the charge. We outline also properties of the generalized null liquid and concrete solutions related to it.

In section \ref{mass_flux}, adopting the results of \cite{Petrov_2021} to the case of 4-dimensional GR, we present the total energy of the aforementioned models and the flux of energy in the form of surface integrals. A structure of such integrals  shows clearly that the total energy can be presented as the energy of the point mass in the Newtonian gravity. One uses more convenient on this stage Eddington-Finkelstein-like ingoing coordinates with the lightlike coordinate $v$ and lightlike sections $v = \const$, like in \cite{Petrov_2021}.

In section \ref{Point_particle}, proposals of previous section are carried out. One uses a reformulation to timelike coordinate $t$ and spacelike sections $t = \const$ that is more convenient for introducing $\del({\bm r})$-function. As a result, in the framework of the field-theoretical derivation we construct the field configurations, which are interpreted as pointlike objects in a space (on a spacelike sections of a background spacetime). After that we present the energy distribution in the $v$ presentation on the lightlike sections $v = \const$ and compare the last with the results in $t$-time presentation.

In section \ref{Concluding}, we outline shortly results and discuss them.

\section{The main notions of the field-theoretical approach}
\setcounter{equation}{0}
\m{ON_FT_approach}

Consider Lagrangian of an arbitrary metric theory:
\be
 \lag(g_{\mu\nu},\Phi^A) = - \frac{1}{2\k}\lag_{g}(g_{\mu\nu}) + \lag_{m}(g_{\mu\nu},\Phi^A)\,,
 \m{lag-g}
 \ee
which depends on the first and second  derivatives of the metric $g_{\mu\nu}$ only. Let the related Euler-Lagrange equations be of the second order, like in GR or in the Lovelock theory; Greek indices numerate spacetime coordinates as $0,1,2,3$. In the present paper, we consider GR in 4 dimensions with the related Lagrangian
\be
\lag_{g}(g_{\mu\nu}) = \sqrt{-g}\l(R - 2\Lambda \r)\,,
 \m{lag_GR}
 \ee
 where $g = \det g_{\mu\nu}$; $\Lambda$ --- the Einstein cosmological constant; the units are chosen in correspondence with $G=c=1$, thus  in (\ref{lag-g}) $\kappa = 8\pi$. In (\ref{lag_GR}), $R$ is the curvature scalar constructed with the use of $g_{\mu\nu}$ from the Ricci tensor $R_{\mu\nu}$: $R = R^\mu_{\mu} = g^{\mu\nu}R_{\mu\nu}$. Then the Einstein equations related to (\ref{lag-g}) with (\ref{lag_GR}) are
\be
R_{\mu\nu} - \half g_{\mu\nu} R+ \Lambda  g_{\mu\nu} = \kappa T_{\mu\nu}\,,
 \m{eqs_GR}
 \ee
where $T_{\mu\nu}$ is the energy-momentum tensor of matter fields $\Phi^A$ with `A' a collective index.

Let the Lagrangian (\ref{lag-g}) and related physical system be considered as perturbed ones with respect to a background system of the Lagrangian $\bar\lag=\lag(\bar g_{\mu\nu},\bar\Phi^A)$, where `bar' means a background quantity. Here, we consider vacuum backgrounds only, for which $\bar\Phi^A=0$. Thus, we consider the background Lagrangian as
\be
\bar\lag=\lag(\bar g_{\mu\nu}, \bar\Phi^A) = - \frac{1}{2\k}\lag_{g}(\bar g_{\mu\nu})= - \frac{1}{2\k}\bar\lag_{g} = - \frac{\sqrt{-\bar g}}{2\k}\l(\bar R - 2\Lambda \r)\,,
 \m{lag_Back}
 \ee
 where $\bar g = \det \bar g_{\mu\nu}$.
In order to transit from the geometrical formulation to the field-theoretical one we decompose variables in (\ref{lag-g}) into
background parts and perturbations\footnote{Here, we apply the decomposition (\ref{varkappa}) used in \cite{Petrov_2019,Petrov_2021} instead of the usual one
$\sqrt{-g}g^{\mu\nu} = \sqrt{-\bar g}(\bar g^{\mu\nu} + h^{\mu\nu})$, see for example, \cite{Petrov+_2017}. Such a different choice does not change the final results, see discussion in \cite{Petrov+_2017}.}:
\bea
 g_{\alf\beta} &=& \bar g_{\alf\beta} + \varkappa_{\alf\beta} \,,
\m{varkappa}\\
\Phi^A &=& \phi^A\,.
\m{phi}
\eea
Metric perturbations are components $\varkappa_{\alf\beta}$, whereas as matter perturbations are $\phi^A$ with respect to a vacuum background spacetime with the metric $\bar g_{\mu\nu}$.
The perturbed system is presented by the {\em dynamical} Lagrangian \cite{GPP,[15]}:
 \be
\lag^{dyn}(\bar g;\,\varkappa,\phi) = \lag (\bar g+\varkappa,\,\phi ) - \varkappa_{\mu\nu} \frac{\delta \bar \lag}{\delta \bar g_{\mu\nu}} - \bar \lag\,,
 \m{lag}
 \ee
where perturbations $\varkappa_{\mu\nu}$ and $\phi^A$ are treated as {\em dynamic variables}.\footnote{The same background metric in (\ref{varkappa}) can be chosen by different ways that leads to different definitions of the field configuration $\varkappa_{\alf\beta}$, where one of them is connected with others by gauge transformations, see, for example, \cite{Petrov_Pitts_2019,Petrov+_2017}.} It is important to note that for small $\varkappa_{\mu\nu}$ and $\phi^A$ expression (\ref{lag}) reduces to the quadratic in {\em dynamic variables} Lagrangian that presents a linearized version of the theory.

After varying the Lagrangian (\ref{lag}) with respect to $\varkappa_{\mu\nu}$ and algebraic transformations one obtains the gravitational equations in the field-theoretical form:
\be
{\cal G}^{L}_{\mu\nu} = \k{\bm t}_{\mu\nu}\,.
 \m{PERTmunu}
 \ee
These equations are equivalent to the field equations of the original theory (\ref{eqs_GR}) if the background equations for $\bar g_{\mu\nu}$:
\be
\bar R_{\mu\nu} - \Lambda \bar g_{\mu\nu} = 0\,
 \m{BACKmunu}
 \ee
are taken into account, see \cite{GPP,[15],Petrov+_2017}. General solutions to the equations (\ref{BACKmunu}) are the Einstein spaces \cite{Petrov_1969}. Among such solutions, following our goals, we consider a flat spacetime and (anti-)de Sitter [(A)dS] spaces, also static Schwarzschild and Schwarzschild-(A)dS black hole spacetimes as background solutions.

The linear operator on the left hand side of (\ref{PERTmunu}) is defined as
 \be
 {\cal G}^{L}_{\mu\nu} =
 \frac{\delta }{\delta \bar g^{\mu\nu}} \varkappa_{\alf\beta}
\frac{\delta \bar \lag_{g}}{\delta \bar { g}_{\alf\beta}}\,.
 \m{GL-q}
 \ee
For the background spacetimes satisfying the equations (\ref{BACKmunu}) one has
 \bea
 {\cal G}^{L}_{\mu\nu} &=&
-\frac{\sqrt{-\bar g}}{2}\l[\bar\nabla_{\rho}{}^{\rho}\varkappa_{\mu\nu} + \bar g_{\mu\nu}\bar\nabla_{\rho\sig}\varkappa^{\rho\sig} - \bar\nabla_{\rho\nu}\varkappa^{\rho}{}_{\mu} - \bar\nabla_{\rho\mu}\varkappa^{\rho}{}_{\nu} - \bar g_{\mu\nu}\bar\nabla_{\rho}{}^{\rho}\varkappa + \bar\nabla_{\mu\nu}\varkappa \r. \nonumber \\  &+& \l. \Lambda \l(2\varkappa_{\mu\nu}- \bar g_{\mu\nu}\varkappa \r)\r]\,,
 \m{GL-kappa}
 \eea
where indices are lowered and raised with the help of the background metric $\bar g_{\mu\nu}$ and $\bar g^{\mu\nu}$; $\varkappa \equiv \varkappa^{\rho}{}_{\rho}$; $\bar\nabla_{\rho}$ is a covariant derivative constructed with the use of $\bar g_{\mu\nu}$; $\bar\nabla_{\rho\sig} \equiv \bar\nabla_{\rho}\bar\nabla_{\sig}$. With regards to the expression (\ref{GL-kappa}), it is worth noting that
 \be
 \bn_\nu {\cal G}_{L}^{\mu\nu}\equiv 0\,,
 \m{div_G_L}
 \ee
if the background equations (\ref{BACKmunu}) hold.

The total symmetric (metric) energy-momentum tensor density for the fields $\varkappa_{\mu\nu}$ and $\phi^A$ on the
right hand side of (\ref{PERTmunu}) is defined in the standard manner:
 \be
 {\bm t}_{\mu\nu} = 2\frac{\delta\lag^{dyn}}{\delta \bar
 g^{\mu\nu}} = {\bm t}^g_{\mu\nu} + {\bm t}^m_{\mu\nu}\,,
 \m{T-q}
 \ee
 where ${\bm t}^g_{\mu\nu} = \sqrt{-\bar g}t^g_{\mu\nu}$  is the energy-momentum density related to a pure gravitational part of the Lagrangian (\ref{lag}), for the explicit complicated expression see \cite{Petrov+_2017}. The energy-momentum density of matter fields $\phi^A$ in (\ref{lag}), ${\bm t}^m_{\mu\nu}$, interacting with $\varkappa_{\alf\beta}$ and propagating in the background spacetime with the metric $\bar { g}_{\alf\beta}$ is connected with the matter energy-momentum in  (\ref{eqs_GR}) by
 \be
 {\bm t}^m_{\mu\nu} = \sqrt{-\bar g}t^m_{\mu\nu} =  \sqrt{-g}T_{\rho\sig}g^{\rho\alf}g^{\sig\beta}\bar g_{\mu\alf}\bar g_{\nu\beta}\,.
 \m{tm_T}
 \ee
  The field equations (\ref{PERTmunu}) with the identity (\ref{div_G_L}) allow us to derive differential conservation law for the total energy-momentum density (\ref{T-q}):
 \be
 \bn_\nu {\bm t}^{\mu\nu} = 0\,.
 \m{div_t}
 \ee

The conservation law (\ref{div_t}) has been derived as a consequence  of the field equations (\ref{PERTmunu}). However, because the Lagrangians noted above are scalar densities, there are more possibilities for deriving conservation laws. We use the diffeomorphism invariance and apply the Noether theorem providing Lie displacements with related an arbitrary enough smooth vector field $\xi^\alf$. After such a procedure interpretation of resulting conserved quantities is left undetermined. It becomes meaningful when vectors $\xi^\alf$ are chosen in a correspondence with a problem under consideration. It could be, for example, Killing vectors of the background spacetime, or proper vectors of observers.

Because the perturbed system is presented by the Lagrangian (\ref{lag}) one can apply the Noether procedure directly to (\ref{lag}). Here, we prefer  a more concise way with economical trick \cite{Petrov+_2017,Petrov_Pitts_2019}, considering an auxiliary Lagrangian:
\be
\lag_1 = -\frac{1}{2\k}\varkappa_{\alf\beta} \frac{\delta \bar
\lag_{g}}{\delta \bar {g}_{\alf\beta}}\,
 \m{Lag-1}
 \ee
that is a part of $\lag^{dyn}$ defined in (\ref{lag}). Because $\lag_1$ is a scalar density one can apply the Noether theorem as well \cite{Petrov+_2017,Petrov_Pitts_2019}. In the result one obtains related identities. After that, making the use of the  field equations (\ref{PERTmunu}), we obtain physically sensible conservation laws
\bea
\di_\mu{\cal I}^{\mu}(\xi) &\equiv & \bn_\mu{\cal I}^{\mu}(\xi) = 0,
\m{CL_current+}\\
{\cal I}^{\mu}(\xi) &= & \di_\nu{\cal I}^{\mu\nu}(\xi)\equiv \bn_\nu{\cal I}^{\mu\nu}(\xi)\,.
\m{CL_super+}
 \eea
 Here, the current ${\cal I}^{\mu}(\xi)=\sqrt{-\bar g }{I}^{\mu}(\xi)$ is a vector density.
 In the case, when the displacement vector is the Killing vector of the background, $\xi^\alf=\bar \xi^\alf$, it becomes
\be
{\cal I}^{\mu}(\bar\xi) = {\bm t}^{\mu}{}_{\nu}\bar\xi^\nu\,.
 \m{current_Lambda}
 \ee
One easily recognizes that substitution  of (\ref{current_Lambda}) into (\ref{CL_current+}) just gives (\ref{div_t}).

The superpotential ${\cal I}^{\mu\nu}(\xi)= \sqrt{-\bar g }{I}^{\mu\nu}(\xi)$ is antisymmetric tensor density, and its components
are calculated as
\be
{\cal I}^{\mu\nu}(\xi) = {{4\over 3}}\l(
 2\xi_\sig \bar\nabla_\lam  {\bm \omega}_{1}^{\sig[\mu|\nu]\lam}  -
{\bm \omega}_{ 1}^{\sig[\mu|\nu]\lam}
\bar \nabla_\lam  \xi_\sig\r)\,,
 \m{C_Law}
\ee
see \cite{Petrov_2019} for arbitrary metric theories with the second order field equations. The quantity ${\bm \omega}_{1}^{\rho\lam|\mu\nu}$ is defined as
\be
 {\bm \omega}^{\rho\lam|\mu\nu}_1  \equiv   \frac{\di \lag_1}{\di \bar g_{\rho\lam,\mu\nu}}\,; \qquad {\bm \omega}^{\rho\lam|\mu\nu}_1  = {\bm \omega}^{\lam\rho|\mu\nu}_1  =
 {\bm \omega}^{\rho\lam|\nu\mu}_1
 \m{NL1}
 \ee
 with $\lag_1$ given in (\ref{Lag-1}); ${\bm \omega}_{1}^{\sig[\mu|\nu]\lam}$ is antisymmetric in $\mu$ and $\nu$. For a concrete theory (\ref{lag_GR}) and (\ref{lag_Back}) with the decomposition (\ref{varkappa}) one has
\be
 {\bm \omega}^{\sig[\mu|\nu]\lam}_1  = \frac{3}{8}\frac{\sqrt{-\bar g}}{\kappa}\l(\bar g{}^{\sig[\mu}\varkappa^{\nu]\lam} - \bar g{}^{\lam[\mu}\varkappa{}^{\nu]\sig} - \varkappa\bar g{}^{\sig[\mu}\bar g{}^{\nu]\lam} \r)\,.
 \m{NL2}
 \ee
Then a superpotential (\ref{C_Law}) acquires the concrete form:
\be
{\cal I}^{\mu\nu}(\xi) = \frac{\sqrt{-\bar g}}{\kappa}\l(\xi^\sig\BN^{[\mu}\varkappa^{\nu]}{}_\sig - \xi^{[\mu}\BN_{\sig}\varkappa^{\nu]\sig} - \varkappa^{\sig[\mu}{\BN}_\sig\xi^{\nu]}
+\xi^{[\mu}{\BN}^{\nu]}\varkappa +\half \varkappa{\BN}^{[\mu}\xi^{\nu]}\r)\,.
 \m{C_Law+}
\ee

Finally, we derive a useful relation for calculating current components. If the field equations  (\ref{PERTmunu}) hold and Killing vectors are used as displacement vectors one has
\be
{\cal I}^{\alf}(\bar\xi) = \sqrt{-\bar g}\l({t}^g_{\mu\nu} + {t}^m_{\mu\nu}\r)\bar g^{\mu\alf}\bar\xi^\nu = \sqrt{-\bar g}\l[\di_\beta I^{\alf\beta}(\bar\xi) + \bar\Gamma^\rho{}_{\rho\beta} I^{\alf\beta}(\bar\xi) \r]\,,
 \m{current_EMT}
 \ee
 where $\bar\Gamma^\alf{}_{\beta\gamma}$ are the Christoffel symbols constructed with the use of $\bar g_{\mu\nu}$.

\section{The Vaidya type solutions }
 \m{Vaidya_type}
\setcounter{equation}{0}

\subsection{The metrics, superpotentials and charges}
 \m{msc}

We study the Vaidya-like solutions, for which a more appropriate form of metric is in the Eddington-Finkelstein generalized coordinates. For our goals we consider the ingoing coordinates only,
\be
ds^2 = -f(r,v)dv^2 +2dvdr + r^2q_{\alf\beta}dx^\alf dx^\beta = -f(r,v)dv^2 +2dvdr + r^2 d\Omega^2
\,,
\m{metric}
\ee
where we follow the presentation of \cite{Cai+_2008}, simplifying solutions to 4 dimensions; as usual, $d\Omega^2 = d\theta^2 +\sin^2\theta d\phi^2$. The advanced null coordinate $v$ is numerated as $v=x^0$, the radial coordinate $r=x^1$ and the angular coordinates $\theta = x^3$ and $\phi = x^3$. We derive also the useful in future expression:
\begin{equation} \label{q_metric}
q_{\alf\beta} = \diag\l(0,0,1,\sin^2\theta\r)\,.
\end{equation}
The solution (\ref{metric}) is considered as perturbed one with respect to a static background solution of the same form:
\be
d\bar s^2 = -\bar f(r)dv^2 +2dvdr + r^2q_{\alf\beta}dx^\alf dx^\beta = -\bar f(r)dv^2 +2dvdr + r^2 d\Omega^2
\,.
\m{metric_bar}
\ee
Thus, the non-zero component of a field configuration propagating on the background (\ref{metric_bar}) and defined by (\ref{varkappa}) is
\be
\varkappa_{00} = g_{00} - \bar g_{00} = - f + \bar f = -\Delta f\,.
\m{varkappa_bar}
\ee

Because our goal is to construct energy and energy flux we choose a displacement vector as $\xi^\alf = \bar\xi^\alf$ in the form of a timelike  Killing vector that for the metric (\ref{metric_bar}) has the general form in all the cases of static $\bar f(r)$:
\be
\bar\xi^\alf = \l\{-1, 0 ,0,0\r\} \,.
\label{bar_xi}
\ee
Then, the non-zero components for the superpotential (\ref{C_Law+}) with (\ref{varkappa_bar}) and (\ref{bar_xi}) are as follows:
\be
{\cal I}^{01}(\bar\xi) = - {\cal I}^{10}(\bar\xi) =
 - \frac{\sin\theta }{\k} r\Delta f .
\label{sup_tau}
\ee
Note that this formula takes a place even in the general case when $\Delta f = \Delta f(v,r)$ with the background metric (\ref{metric_bar}).

All of these allows us to construct conserved gravitational charges inside a 2-surface $\di\Sigma$ by a surface integration:
\be
\l.{\cal P}(\bar\xi)\r|_{\di\Sigma} = \oint_{\di\Sigma}dx^{2}{\cal I}^{01}(\bar\xi)\,.
\m{charge}
\ee
Here, $\di\Sigma$ is defined by intersection of a hypersurface defined by $r= \const$ with a light-like section $\Sigma := v = \const$.
To construct a flux of the quantity (\ref{charge}) through $\di\Sigma$ one has to differentiate it with respect to $v$ that is denoted by dot:
\be
\l.\dot{\cal P}(\bar\xi)\r|_{\di\Sigma} = \oint_{\di\Sigma}dx^{2}\di_0{\cal I}^{01}(\bar\xi)  = - \oint_{\di\Sigma}dx^{2}{\cal I}^{1}(\bar\xi)
\,.
\m{flux}
\ee
It is explaned by the definition (\ref{CL_super+}), ${\cal I}^{1} = \di_0{\cal I}^{10}$. Thus, the quantity ${\cal I}^{1}$ presents the flux density of the quantity ${\cal P}(\xi)$ through the surface $\di\Sigma$. The minus sign in (\ref{flux}) is interpreted by the fact that the flux is opposite to the direction of the coordinate $r$.

\subsection{The Vaidya-like solutions with a generalized null liquid. }
 \m{lightlike_matter}

To derive the solution of the Vaidya type with the generalized null liquid we follow the work \cite{Cai+_2008} and simplify its results to 4 dimensions. Let us analyze the equations (\ref{eqs_GR}) presented by the metric of the Vaidya form in ingoing coordinates (\ref{metric}). From the beginning we assume that the energy-momentum for the null liquid is presented by the component $T_0{}^1=\mu$ in (\ref{eqs_GR}) only, the other components $T_\alf{}^\beta$ are equalized to zero. Then, integrating in $r$ a combination of the components $\{_0{}^0\}$ and $\{_i{}^k\}$ of the equations (\ref{eqs_GR}), we introduce into consideration $M(v)$ that is an arbitrary function of $v$. It gives the simplest and standard null liquid presentation.

To introduce the {\em generalized} energy-momentum for null liquid we follow \cite{Cai+_2008} and \cite{Dominguez_Gallo_2006}. Let us assume $T_0{}^0 =T_1{}^1 \neq 0$ together with $T_0{}^1 \neq 0$. After that one integrates  the consistency condition $\nabla_\beta T_{\alf}{}^{\beta} = 0$ and obtain  $T_0{}^0 =T_1{}^1 = {C}(v)r^{2\sigma-2}/\kappa$ with an arbitrary constant $\sigma$, where $\nabla_\beta$ is a covariant derivative with respect to $g_{\mu\nu}$, and ${ C}(v)$ is another arbitrary function of $v$. The inevitable interpretation of non-zero $T_1{}^1$ is a radial pressure $P_r = - T_1{}^1$. Thus, non-zero ${ C}(v)$ signals the presence of $P_r$ and defines its time evolution. It is added a tangential pressure $P_t = -\sigma P_r$ as well.

Finally, the generalized energy-momentum for the null liquid becomes
\be
{T}_{\alf\beta} = \mu\l( n_\alf n_\beta \r)+ P_r \l(n_\alf l_\beta + l_\alf n_\beta \r) +  r^2P_t\,q_{\alf\beta}\,.
\m{T_light}
\ee
Ingoing and outgoing null vectors are $n^\alf = \l(0,-1, 0,0\r)$ and $l^\alf = \l(1,f/2, 0, 0\r)$, respectively,
for which $n^\alf n_\alf = 0$, $l^\alf l_\alf = 0$, $l^\alf n_\alf = -1$; $q_{\alf\beta}$ is defined in (\ref{q_metric}). The energy density and the radial pressure
\bea
 \mu &=& \frac{1}{\k}\l( \frac{2\dot M(v)}{r^2} + \frac{\dot { C}(v)\Theta (r) }{r^2}\r)\,,
\m{mu_light}\\
P_r &=& - \frac{1}{\k}{ C}(v)\frac{\Theta'(r)}{r^2}    = -\frac{1}{\k}{ C}(v)r^{2\sigma-2}\,
\m{Pr_light}
\eea
are obtained by the requirement of the consistency of the system of the equations (\ref{eqs_GR}). An appearance of $\Theta (r)$ is a result of integration when $\mu$ is constructed.
The dominant energy conditions applied in \cite{Dominguez_Gallo_2006} to the energy-momentum (\ref{T_light}) presenting the null fluid of the type $II$ in classification of \cite{Hawking_Ellis_1973} give
$\mu \geq 0$,  ${\cal C}(v) \le 0$ and $-1 \le \sig \le 0$. More details are
\bea
&&\sig = -1/2 ~~\goto ~~ \Theta(r) = \ln(r)\,;
\m{Theta_1}\\
&&-1 \le \sig \le 0 ~~ {\rm with}~~\sig \neq -1/2  ~~\goto ~~  \Theta(r) = \frac{r^{2\sig +1}}{2\sig +1}\,.
\m{Theta_2}
\eea

Finally, the equations (\ref{eqs_GR}) are resolved  as
\be
f(r,v) = 1 - \l[\frac{r^2}{\ell^2} + \frac{ 2M_0 + 2M(v)}{r} + \frac{{C}(v)\Theta (r)}{r }\r]
\,,
\m{odd+}
\ee
where $\ell^2 \equiv 3/\Lambda$.
In the case $M(v) = {C}(v) = 0$ the solution (\ref{odd+}) is simplified to the usual solutions for static Schwarzschild-like black holes with the mass parameter $M_0$. The cases $1/\ell^2<0$, $1/\ell^2>0$ and $1/\ell^2=0$ correspond to solutions with AdS, dS and flat asymptotics, respectively.


It is natural to restrict ourselves by the following simple requirements: 1) the constant mass parameter, defining static black holes, satisfies $M_0>0$; 2) a presence of radiating matter is defined by $\mu>0$; 3) then, from (\ref{mu_light}), it follows that the sum $ 2M(v) + {C}(v)\Theta (r)$ is increasing with $v$ and, thus,  corresponds to {\em monotonic evolution}. We do not consider $M(v)$ and ${C}(v)$ separately, as this is sufficient for purposes of the present paper. Various variants of relations between $M(v)$ and ${\cal C}(v)$ in the Einstein-Gauss-Bonnet gravity have been considered in \cite{Dominguez_Gallo_2006}.

\section{Total masses and their fluxes}
 \m{mass_flux}
\setcounter{equation}{0}

\subsection{Static solutions}
 \m{Static}

From the start let us consider the general solution (\ref{odd+}) in static regime:
\be
f(r) = 1 - \frac{r^2}{\ell^2} - \frac{2M_0 }{r}
\,
\m{static_tot}
\ee
for all the three cases of the cosmological constant: $1/\ell^2=0$, $1/\ell^2<0$ and $1/\ell^2>0$.

In the case of {\em flat asymptotics}, $1/\ell^2=0$, one has the standard Schwarzschild solution in the ingoing Edington-Finkelstein coordinates:
\be
f(r) = 1 -\frac{2M_0 }{r}
\,,
\m{static_flat}
\ee
that presents the static black hole with the only one event horizon  $r_+$ satisfying $f(r_+)=0$ that gives $r_+ = 2M_0$.

For the {\em AdS asymptotics}, when  $1/\ell^2<0$ the static solution (\ref{odd+}) becomes
\be
f(r) = 1+ \frac{r^2}{\l|\ell^2\r|} - \frac{ 2M_0}{r}
\,.
\m{static_AdS}
\ee
This presents a static black hole (Schwarzschild-AdS black hole) with the only one event horizon $r_+$ satisfying $f(r_+)=0$ that gives
\be
1+\frac{r_+^{2}}{\l|\ell^2\r|}  =  \frac{ 2M_0}{r_+}
\,.
\m{horizon_AdS}
\ee

The case with the {\em dS asymptotics}, when  $1/\ell^2>0$, is a more complicated one. The static solution (\ref{odd+}) becomes
\be
f(r) = 1 -  \frac{r^2}{\l|\ell^2\r|} - \frac{ 2M_0}{r}
\m{static_dS}
\ee
that presents the Schwarzschild-dS black hole. Recall that dS space has the cosmological  horizon $r_c = \ell$. Keeping this fact in mind, one easily concludes that for the enough small $M_0$ there are the two horizons: the cosmological one and the event horizon. They satisfy the equation:
\be
1-\frac{r_{c,+}^{2}}{\l|\ell^2\r|}  =  \frac{ 2M_0}{r_{c,+}}
\,.
\m{horizon_dS}
\ee
When $M_0$ becomes $M_{cr} = \ell/(\sqrt{3})^3$, then the horizons $r_+$ and $r_c$ are united to $r_+ = r_c = \ell/\sqrt{3}$ and the spacetime becomes the black hole of the Nariai type \cite{Nariai_1950,Nariai_1951}, where properties of a spacetime outside the horizon with respect to properties of a spacetime inside the horizon are unusual. When $M_0> M_{cr}$ one finds a naked singularity at $r=0$ without horizons. But such an interpretation is unusual because now $r=0$ presents a spacelike surface, unlike for an ordinary naked singularity (say, in the  Raissner-Nordstr\"om solution with correspondent parameters), where $r=0$ presents a timelike world line.

Each of the asymptotic spacetimes can be chosen as a background spacetime:
\be
\bar f(r) = 1 -  \frac{r^2}{\ell^2}
\,.
\m{static_bar}
\ee
for all the three cases of $1/\ell^2$. Then independently on the choice of $1/\ell^2$ one has the unified form for the perturbations:
\be
\Delta f = f(r) - \bar f(r) = - \frac{ 2M_0}{r}
\,.
\m{static_pert}
\ee
To calculate the energy (mass) of static black hole we choose the timelike Killing vector (\ref{bar_xi}). Then the superpotential component (\ref{sup_tau}) acquires the very simple form:
\be
{\cal I}^{01}(\bar\xi) = M_0\frac{ \sin\theta}{4\pi}
\,
\m{static_sup}
\ee
 that is the same in all the cases $1/\ell^2$.
Substituting it into (\ref{charge}) and choosing the boundary of integration as a sphere of the radius $r_0$, we obtain
\be
\l.{\cal P}(\bar\xi)\r|_{r_0}= \oint_{r_0}dx^2{\cal I}^{01}(\bar\xi) = M_0\oint_{r_0}d\theta d\phi\frac{ \sin\theta}{4\pi} = M_0
\,.
\m{static_mass}
\ee

Let us discuss this result. First, it is the acceptable result $M_0$ that takes a place for all the three cases. Second, the result (\ref{static_mass}) does not depend on $r_0$ that means that in all the cases the total mass of configuration (\ref{static_pert}) is concentrated at the point $r=0$ of the background. Thus this presentation could be interpreted as a point particle that is studied in detail in the next section.

We recall that dS black holes have unusual properties. Thus, in \cite{dS_1,dS_2} authors have assumed and proved that if static black holes placed in dS space, then the total masses of the models are less than the mass of the initial non-singular dS space without a black hole. This result qualitatively coincides in our conclusion in \cite{Petrov_2021} where we calculate a quasi-local energy in $D>4$ for perturbations presenting Schwarzschild-dS black hole on the dS background. Here, the result (\ref{static_mass}) in $D=4$ has another interpretation, it is only as a mass point concentrated at $r=0$. It takes a place even for $M_0 \geq M_{cr}$, when the solution is interpreted as the Nariai black hole or a naked singularity.
In all the values of $M_0$ we consider the unique dS background, and a consideration is restricted by the cosmological horizon of the background.

\subsection{Dynamic solutions on the symmetric backgrounds}
 \m{Dynamic}

Let us turn to the total solution (\ref{odd+}), preserving the time dependence on $v$.
Again, from the start, we choose the asymptotic spacetimes as a background spacetime, like in (\ref{static_bar}). Then the field configuration is described by the perturbation:
\be
\Delta f = f(v,r) - \bar f(r) = - \frac{ 2M_0 +2M(v)}{r} -\frac{ C(v)\Theta(r)}{r}
\,.
\m{dynamic_pert}
\ee
In the case of {\em absence of the pressure} for the null liquid $C(v) = 0$ we can repeat all the steps in the static case. Now, in spite of dependence on $v$ superpotential (\ref{sup_tau}) gives the analogical to (\ref{static_sup}) expression:
\be
{\cal I}^{01}(\bar\xi) = \l[M_0 + M(v)\r]\frac{ \sin\theta}{4\pi}
\,,
\m{dynamic_sup}
\ee
and for the energy of the object
\be
\l.{\cal P}(\bar\xi)\r|_{r_0}=  M_0 + M(v)
\,.
\m{dynamic_mass}
\ee

Interpretation of (\ref{dynamic_mass}) is almost analogical to the interpretation of (\ref{static_mass}). First, it is the acceptable result $M_0 + M(v)$, it takes a place for all the three cases.
Second, the result (\ref{dynamic_mass}) does not depend on $r_0$ that means that in all the three cases the total mass of configuration (\ref{static_pert}) is concentrated at the point $r=0$ of the background. It is surprisingly because the space is filled by the null liquid that has its own energetic characteristics. All of these properties is studied in detail in the next section.
The energy (\ref{dynamic_mass}) is rising with $v$ due to the starting assumptions. It can be supported by the formula (\ref{flux}):
\be
\l.\dot{\cal P}(\bar\xi)\r|_{r_0}=  \dot M(v) >0
\,.
\m{dynamic_flux}
\ee

Concerning the dS case, let us consider the evolution from the moment when $M=M_0:= 0<M_0<M_{cr}$. Then, due to (\ref{dynamic_flux}) at the moment $v=v_{cr}$ mass reaches the value
$M=M_0 + M(v_{cr})$ and the Schwarzschild-dS black hole becomes the Nariai black hole. After that it becomes the aforementioned naked singularity, $M>M_{cr}$. Nevertheless, at all the stages the result (\ref{dynamic_mass}), that signals on a concentration of the energy at the point $r=0$, is preserved.

Now, let us switch on the pressure of the null liquid and consider the perturbation (\ref{dynamic_pert}) in whole $C(v)\neq 0$. Then the superpotential acquires a more complicated form:
\be
{\cal I}^{01}(\bar\xi) = \l[M_0 + M(v) + \frac{C(v)\Theta(r)}{2}\r]\frac{ \sin\theta}{4\pi}
\,,
\m{dynamic_sup_C}
\ee
that depends on $r$ unlike the previous expressions. Its integration gives the  energy inside the sphere $r = r_0$
\be
\l.{\cal P}(\bar\xi)\r|_{r_0}=   M_0 + M(v)+\frac{C(v)\Theta(r_0)}{2}
\,.
\m{dynamic_mass_C}
\ee
Here, one can see that, unlike (\ref{dynamic_mass}), not all the energy is concentrated at the center. Only the energy defined by $M_0 + M(v)$  can be imitated by the point particle distribution. Really, the expression (\ref{dynamic_mass_C}) is a quasilocal expression. We will discuss this situation in the next section.
By the starting assumption the quantity (\ref{dynamic_mass_C}) is rising with $v$, indeed, due to (\ref{flux}) one has
\be
\l.\dot{\cal P}(\bar\xi)\r|_{r_0}=  \dot M(v) +\frac{\dot C(v)\Theta(r_0)}{2} >0
\,.
\m{dynamic_flux_С}
\ee

In the case of flat and AdS asymptotics one can find the global charges setting $r_0 \goto \infty$. In this case we check the behaviour of $\Theta(r)$ presented in (\ref{Theta_1}) and (\ref{Theta_2}). It is evidently that (\ref{Theta_1}) leads to an infinity value of (\ref{dynamic_mass_C}) that is not reasonable physically. Moreover, such a situation contradicts to the requirement (ii) in Introduction.  Therefore, the behaviour in (\ref{Theta_1}) and (\ref{Theta_2}) is restricted to
\be
-1 \le \sig <  -1/2 ~~ \goto ~~  \Theta(r) = \frac{r^{2\sig +1}}{2\sig +1}\,
\m{Theta_2+}
\ee
only. Then for the flat and AdS asymptotics one has
\be
\l.{\cal P}(\bar\xi)\r|_\infty=   M_0 + M(v)
\,.
\m{dynamic_mass_inf}
\ee
However, for finite $r_0$, of course $\l.{\cal P}(\bar\xi)\r|_{r_0}\neq\l.{\cal P}(\bar\xi)\r|_\infty$.

In the case of the dS asymptotics we cannot to provide the limit $r_0 \goto \infty$ because the  dS background has the cosmological horizon and it is not reasonable to go behind of that. Thus, in the dS case, for the quasi-local expression (\ref{dynamic_mass_C}) one can use the more soft behaviour (\ref{Theta_1}) and (\ref{Theta_2}).

\subsection{Dynamic solutions on the black hole backgrounds}
 \m{Dynamic_on_BH}

Because the field-theoretical formalism is an universal one we can choose the geometries of static black holes as backgrounds for perturbations conditioned by the null liquid falling into these black holes. Because the solutions for the static black holes are vacuum ones we can use the general formulae of the section \ref{ON_FT_approach}. Thus, we choose
\be
\bar f(r) = 1 -  \frac{r^2}{\ell^2} - \frac{2M_0}{r}
\,
\m{bar_BH}
\ee
as a background solution for all the cases $1/\ell^2$. In the dS case, we consider the black hole as well, that is we restrict ourselves by $0<M_0<M_{cr}$. Then independently on a choice of $1/\ell^2$ one has the only one form for the perturbations:
\be
\Delta f = f(v,r) - \bar f(r) = - \frac{ 2M(v)}{r} -\frac{ C(v)\Theta(r)}{r}
\,.
\m{BH_pert}
\ee
Again, to calculate the energy (mass) of perturbations on the background of static black holes we can choose the timelike Killing vector in the same form (\ref{bar_xi}). Then the superpotential (\ref{sup_tau}) acquires the form:
\be
{\cal I}^{01} = \l[M(v) +  \frac{ C(v)\Theta(r)}{2} \r]\frac{ \sin\theta}{4\pi}
\,.
\m{BH_sup}
\ee

Then, in the case of absence of the pressure of null liquid $C(v)=0$ we obtain that the energy of the object becomes:
\be
\l.{\cal P}(\bar\xi)\r|_{r_0}=   M(v)
\,,
\m{BH_mass}
\ee
where we keep in mind $r_0 > r_+$. Thus we consider the boundary of integration outside the horizon $r_+$ of the static background black hole in all the three  cases. Therefore, it is impossible to interpret the result (\ref{BH_mass}) as the concentration of the  energy at a point. However, at least, we can interpret it by the way that the total energy of all the perturbations is concentrated inside the horizons of the background black holes. This situation is close to the model considered in \cite{Lundgren_Schmekel_York_2007} where, basing on the Brown-York approach \cite{Brown_York_1993} the authors interpret the energy of the black hole inside the horizon without forming a point mass. In fact, the result (\ref{BH_mass}) presents the energy collected by the falling null liquid during the time of the falling.

Including the pressure of the null liquid $C(v)\neq 0$ into consideration, we obtain the related to (\ref{dynamic_sup_C}) - (\ref{dynamic_mass_inf}) formulae with excluding $M_0$ only and with the totally similar interpretation. Especially, we note that restrictions to (\ref{Theta_2+}) are left the same in the case of the background Schwarzschild and Schwarzschild-AdS black holes.

However, because the results of this subsection cannot be interpreted with making the use of the pointlike representation we will not develop them further.

\section{Point particle interpretation}
 \m{Point_particle}
\setcounter{equation}{0}

In this section, we develop the announced above proposal that in the field-theoretical derivation the field configuration for the general solution (\ref{odd+}) can be interpreted as a massive point particle described by $\del({\bm r})$-function considered in a background spacetime. From the start, following this goal we transform the presented above formalism to timelike coordinate $t$ and spacelike sections $t = \const$. What are reasons to such a transformation? First, we study the energy distribution that just has a natural description on spacelike hypersurfaces. One could define spacelike sections in null $v$ coordinates, but then one needs operate with very cumbersome expressions. Later, for the sake of brevity and uniform we will call $v$ it as `time' as well. Second, the initial definition of $\del({\bm r})$-function is introduced in 3-dimensional space that can be interpreted as a spacelike section. Third, the $t$-time presentation is very convenient for a comparison with earlier works. Moreover, for the best of our knowledge, there is no works in a pointlike description in $v$-time presentation. Fourth, it is interesting to outline properties of energy density of the null liquid just under the $t$-time presentation, although, in the field-theoretical formalism. Indeed, usually these properties are given in the Eddington-Finkelstein-like coordinates.

Nevertheless, because the field-theoretical formalism is a covariant one we anticipate that a pointlike description in the $v$-time presentation on lightlike sections takes a place as well. Although it looks as a formal description, we demonstrate it.

\subsection{Charges and fluxes of charge in the $t$-time presentation}
 \m{t_presentation}

Thus, we represent the metric of the background spacetime (\ref{metric_bar}) and the related field configuration (\ref{varkappa_bar}) in the standard time coordinate $t$ instead of the null coordinate $v$. We use the transformation
\be
dv=dt+\frac{dr}{\bar f(r)}
\,,
\m{v_t}
\ee
preserving the other coordinates: $r$, $\theta$ and $\phi$. Ignoring the integration constants it can be rewritten in the form of an indefinite integral:
\be
v(t,r)=t+\int^r\frac{d\rho}{\bar f(\rho)} \equiv t+\psi(r)\,.
\m{v_t_int}
\ee
In the result, the metric (\ref{metric_bar}) transforms to
\be
d\bar s^2 =-{\bar f(r)}dt^2 + \frac{dr^2}{{\bar f(r)}}+ r^2d\Omega^2
\,,
\m{metric_t}
\ee
for which the Christoffel symbols are
\bea
&&\bar \Gamma^0{}_{01} = - \bar \Gamma^1{}_{11} = \frac{\bar f'}{2\bar f}\,,~~\bar \Gamma^1{}_{00}=\half{\bar f'}{\bar f}\,,~~\bar \Gamma^2{}_{12}=\bar \Gamma^3{}_{13}=\frac{1}{r},\nonumber\\
&&\bar \Gamma^1{}_{22}=  -r\bar f,~~\bar \Gamma^1{}_{33}=-r\bar f\sin^2\theta,~~\bar \Gamma^2{}_{33}=-\sin\theta\cos\theta,~~ \bar \Gamma^3{}_{23}=\cot\theta
\,;
\m{Gamma_t}
\eea
now and below the coordinate $t$ is numerated by the index `0'; indices are raised and lowered with the use of the metric (\ref{metric_t}).
The field configuration (\ref{varkappa_bar}) transforms to
\be
\varkappa_{00}= -\Delta f,\qquad \varkappa_{01}= -  \frac{\Delta f}{{\bar f(r)}}, \qquad \varkappa_{11}= -  \frac{\Delta f}{{\bar f^2(r)}}
\,.
\m{varkappa_t}
\ee
Unlike (\ref{dynamic_pert}), the perturbations have to be considered with
\be
\Delta f =  - \frac{ 2M_0 +2M\l[v(t,r)\r]}{r} -\frac{ C\l[v(t,r)\r]\Theta(r)}{r}
\,
\m{dynamic_pert_t}
\ee
that is with evident dependence on $t$ and $r$.

By the transformation (\ref{v_t}) the form of the Killing vector (\ref{bar_xi}) and of the non-zero components of the superpotential (\ref{sup_tau}) in coordinates (\ref{metric_t}) are left the same, with the $t$-time interpretation of the index `0'. Thus, for the non-zero component of the superpotential one has
\be
{\cal I}^{01}(\bar\xi) = - {\cal I}^{10}(\bar\xi) = \sqrt{-\bar g}{I}^{01}(\bar\xi) =
  \frac{\bar\xi^0 \sqrt{-\bar g}}{\k} \frac{\Delta f}{r}
\,,
\m{sup_t}
\ee
where $\Delta f = \Delta f\l[v(t,r),r\r] $ is defined in (\ref{dynamic_pert_t}). Of course, the same quantity is obtained by a direct substitution of (\ref{metric_t})-(\ref{varkappa_t}) into (\ref{C_Law+}). Analogously to (\ref{charge}) one constructs the conserved charge inside the surface ${\di\Sigma}$
\be
\l.{\cal P}(\bar\xi)\r|_{\di\Sigma} = \oint_{\di\Sigma}dx^{2}{\cal I}^{01}(\bar\xi)\,.
\m{charge_t}
\ee
Only now, ${\di\Sigma}$ is defined by intersection of hypersurface $r = \const$ with the timelike section ${\Sigma} := t = \const$.

One easily recognizes that really the surface ${\di\Sigma}$ is the same one in both the cases of $v$- and $t$-presentations. However, due to the transformations (\ref{v_t}) and (\ref{v_t_int}) one has a dependence on $r_0$ in the charge (\ref{charge_t}) in general, if $\di\Sigma$ is presented by a sphere of radius $r_0$. Thus, let us step by step analyze the expression (\ref{charge_t}). From the beginning, let us restrit ourselves to the static case. Then, we obtain the result coinciding exactly with the charge (\ref{static_mass}) in the $v$-presentation:
\be
\l.{\cal P}(\bar\xi)\r|_{r_0} = \oint_{r_0}dx^{2}{\cal I}^{01}(\bar\xi) = M_0\,,
\m{charge_static}
\ee
that does not depend on $r_0$ anyway. Thus, we see again that all the energy is concentrated at the point $r=0$ on $\Sigma$.

Another picture is when we switch on the null liquid. In the case of absence  of the pressure we have
\be
\l.{\cal P}(\bar\xi)\r|_{r_0} = \oint_{r_0}dx^{2}{\cal I}^{01}(\bar\xi) = M_0 +M[v(t,r_0)]= M_0 +M[t+\psi(r_0)]\,
\m{charge_Mv}
\ee
instead of (\ref{dynamic_mass}). Therefore, now there is no a concentration of the energy totally at the center. It is not surprisingly because at a less values $r_0$ we observe a less quantity of null liquid under the surface $r=r_0$ on the hypersurface $\Sigma$. At last, let us include into consideration the pressure of the null liquid. Then, instead of (\ref{dynamic_mass_C}) one has
\bea
\l.{\cal P}(\bar\xi)\r|_{r_0}& =& \oint_{r_0}dx^{2}{\cal I}^{01}(\bar\xi) = M_0 +M[v(t,r_0)] +\frac{C[v(t,r_0)]\Theta(r_0)}{2}\nonumber\\
& =& \oint_{r_0}dx^{2}{\cal I}^{01}(\bar\xi) = M_0 +M[t+\psi(r_0)] +\frac{C[t+\psi(r_0)]\Theta(r_0)}{2}\,.
\m{charge_MC}
\eea
Interpretation of this expression is close to that of the expression (\ref{charge_Mv}). Below, all the results (\ref{charge_static})-(\ref{charge_MC}) will be explained by derivation of the energy distribution (energy density) on the spacelike section $\Sigma:=t=\const$.

A consideration of fluxes in the present section where we search for the pointlike mass distribution is not so appropriate. Nevertheless, lets us give related notions in the $t$ presentation. The flux of the charge (\ref{charge_t}) through the surface $\di\Sigma$ is defined by its differentiation with respect to $t$:
\be
\frac{d}{dt}\l.{\cal P}(\bar\xi)\r|_{r_0} = \frac{d}{dt}\oint_{r_0}dx^{2}{\cal I}^{01}(\bar\xi) = - \oint_{r_0}dx^{2}{\cal I}^{1}(\bar\xi)\,,
\m{flux_t}
\ee
where ${\cal I}^{1}$ is defined as usual by (\ref{CL_super+}), but in the $t$ presentation now. Comparing ${\cal I}^{1}$ in (\ref{flux}) and ${\cal I}^{1}$ in (\ref{flux_t}) one finds that it is the same component with the exchange of the dependence by (\ref{v_t_int}). Thus, all the conclusions concerning fluxes in $t$-presentation and in  $v$-presentation coincide.

\subsection{The pointlike energy distribution. The $t$-time presentation}
 \m{point}

Now, let us recall the necessary formulae related to the definition of the Dirac $\delta$-function, see \cite{GSh_1964,Korn_1969}. Firstly, it is the simplest formula:
\be
\int_{\Sigma} dx^3 \chi({\bm r})\delta({\bm r})  = \chi(0) \,.
\m{int_delta_c}
\ee
The spacelike section again is defined as $\Sigma:=t=\const$, and $\chi({\bm r})$ is a scalar. One has to accent that formula (\ref{int_delta_c}) is the formal one and requires a separate interpretation in any concrete case. It is quite evident for the case when  $\Sigma$ presents a flat space and Cartesian coordinates $(x,y,z)$ are used with $r^2=x^2+y^2+z^2$.
Let us assume that the spherical coordinates $(r,\theta,\phi)$  are connected with $(x,y,z)$ in an usual way. Then a reformulation in (\ref{int_delta_c}) has to be made as
\be
\int_{\Sigma} dx^3 \sqrt{\bar g_3}\chi({\bm r})\delta({\bm r})  = \chi(0) \,,
\m{int_delta}
\ee
where $\sqrt{\bar g_3} = r^2\sin\theta$. The formula (\ref{int_delta}) could be used for curved spaces with a spherical symmetry as well, see, for example, \cite{Spirin_2022}. Another useful formal  equality is
\be
\nabla^2 \l(\frac{1}{r} \r) = \l(\frac{d^2}{dr^2} + \frac{2}{r} \frac{d}{dr}\r)\frac{1}{r} = -4\pi\delta(\bm r) \,,
\m{delta}
\ee
where $\nabla^2$ is the Laplace operator.

To obtain the  mass distribution (energy density) of the perturbations (\ref{varkappa_t}) on the hypersurface ${\Sigma} := t = \const$ one has to calculate
 the `0'-th component of the current ${\cal I}^0$ with the Killing vector $\bar\xi^\alf = (-1,0,0,0)$. We turn to the relation (\ref{current_EMT}) and use the concrete expression (\ref{sup_t}) for the superpotential components with (\ref{dynamic_pert_t}). Thus,
\be
{ I}^{01}(\bar\xi) = -\frac{1}{\kappa}\frac{d}{dr}\l(\frac{1}{r} \r)\l\{2M_0 +2M[v(t,r)] +C[v(t,r)]\Theta(r)\r\}\,.
\m{I_01}
\ee
Differentiating it one has
\bea
\di_1{ I}^{01}(\bar\xi) = &&-\frac{1}{\kappa}\frac{d^2}{dr^2}\l(\frac{1}{r} \r)\l\{2M_0 +2M[v(t,r)] +C[v(t,r)]\Theta(r)\r\} \nonumber\\
&&+ \frac{1}{\kappa r^2}\l\{2{\dot M[v(t,r)]}\frac{1}{\bar f} +{\dot C[v(t,r)]}\frac{\Theta(r)}{\bar f} +{C[v(t,r)]\Theta'(r)}\r\}\,,
\m{diI_01}
\eea
where it was used the relation (\ref{v_t_int}); `dot' again means a differentiation with respect to $v$, and `prime'  means a differentiation with respect to $r$.

Substituting (\ref{I_01}) and (\ref{diI_01}) into (\ref{current_EMT}) and taking into account (\ref{mu_light}), (\ref{Pr_light}) and (\ref{delta}), one finds
\bea
{\cal I}^{0}(\bar\xi) &=&\sqrt{-\bar g}t^0{}_0\bar\xi^0= \sqrt{-\bar g}\l\{M_0 +M[v(t,r)] +\frac{C[v(t,r)]\Theta(r)}{2}\r\} \delta(\bm r)\nonumber\\
&+& \frac{\sqrt{-\bar g}}{\bar f}\l\{ \mu[v(t,r),r] -  P_r[v(t,r),r]\bar f
\r\}\,.
\m{I_0}
\eea
This expression presents the energy distribution for the model under consideration with the generalized null liquid.

From the beginning we consider the static case when the null liquid is absent: $M(v)=C(v)=0$. Then (\ref{I_0}) can be rewritten as
\be
{\cal I}^{0}(\bar\xi) =\sqrt{-\bar g}t^0{}_0\bar\xi^0= \sqrt{-\bar g}M_0 \delta(\bm r)\,.
\m{I_0_static}
\ee
The volume integration of this expression under these conditions gives
\be
\l.{\cal P}(\bar\xi)\r|_{r_0} = \int^{r_0}_0 dx^{3}{\cal I}^{0}(\bar\xi)= \int^{r_0}_0 dx^{3}\sqrt{-\bar g}M_0 \delta(\bm r) = M_0\,
\m{static_M0}
\ee
that has been obtained with the use of  (\ref{int_delta}) because here  $\sqrt{-\bar g} = \sqrt{\bar g_3}$. First, the result (\ref{static_M0}) repeats the result  (\ref{charge_static}) that tells us on an absence of contradictions in our constructions. Second, the result (\ref{static_M0}) tells us that, indeed, total energy density of gravitational field without null liquid is described by the pointlike distribution with the use of the $\delta$-function:
\be
t^g_{00} = M_0 \bar f(r)\delta(\bm r) \,,
\m{t_00_g}
\ee
see (\ref{T-q}), where $t^m_{00}=0$.
Third, the results (\ref{static_M0}) and (\ref{t_00_g}) hold in all the cases of the background: flat and (A)dS. Thus, the result (\ref{static_M0}) generalises to the (A)dS asymptotics the result in \cite{Petrov_2005,Petrov_2018,Petrov+_2017}, where the Schwarzschild black hole with the flat asymptotics, $\bar f = 1$, only was represented in the form of mass point particle.

Considering the dynamic case with the null liquid, but without pressure $C(v)=0$, we obtain from (\ref{I_0}):
\be
{\cal I}^{0}(\bar\xi) =\sqrt{-\bar g}t^0{}_0\bar\xi^0= \sqrt{-\bar g}\l\{M_0 + M[v(t,r)]\r\}\delta(\bm r) +\frac{\sqrt{-\bar g}}{\bar f} \mu[v(t,r),r]\,.
\m{I_0_dinamicM}
\ee
The volume integration of this expression  gives
\be
\l.{\cal P}(\bar\xi)\r|_{r_0} = \l.\int^{r_0}_0 dx^{3}{\cal I}^{0}(\bar\xi)=M_0+ M[v(t,0)] + M[v(t,r)]\r|^{r_0}_0=M_0 + M[v(t,r_0)]\,.
\m{dynamicMv}
\ee
Again the result (\ref{dynamicMv}) has been obtained with the use of  (\ref{int_delta}). First, the result (\ref{dynamicMv}) repeats the result  (\ref{charge_Mv}) that tells us on an absence of contradictions in our constructions. Second, turning to (\ref{T-q}) we try to distinguish energy distribution (\ref{I_0_dinamicM}) into gravitational and matter parts. Using (\ref{T_light}) in (\ref{tm_T}) in coordinates (\ref{metric_t}) we easily find that $t^m_{00} =\mu[v(t,r)]$. Then, for $t_{00} =t^g_{00}+t^m_{00}$ in (\ref{I_0_dinamicM}) we have
\bea
t^g_{00} &=& \l\{M_0 +  M[v(t,r)]\r\}\bar f(r)\delta(\bm r) \,.
\m{t_00_g+}\\
t^m_{00} &=&  \mu[v(t,r),r]\,.
\m{t_00_m+}
\eea
It is interesting to note that the energy distribution outside $r=0$ is defined by the null liquid only, without the accompanying energy density of gravitational field!  Then,
the energy density of the pure gravitational field is described by the pointlike distribution with the use of the $\delta$-function (\ref{t_00_g+}), although with influence of the matter field.
Third, the results (\ref{dynamicMv})-(\ref{t_00_m+}) hold in all the cases of the background: flat and (A)dS.

At last, let us turn to the expression for the energy density (\ref{I_0}) in whole. Its volume integration gives
\bea
\l.{\cal P}(\bar\xi)\r|_{r_0} &=& \int^{r_0}_0 dx^{3}{\cal I}^{0}(\bar\xi)
= M_0+ M[v(t,0)] +\frac{C[v(t,0)]\Theta(0)}{2}\m{dynamicMC}\\ &+& \l.\l\{M[v(t,r)] +\frac{C[v(t,r)]\Theta(r)}{2}\r\}\r|^{r_0}_0=M_0 + M[v(t,r_0)]+\frac{C[v(t,r_0)]\Theta(r_0)}{2}\,.
\nonumber
\eea
The result (\ref{dynamicMC}) has been obtained with the use of  (\ref{int_delta}). First, the result (\ref{dynamicMC}) repeats the result  (\ref{charge_MC}) that tells us on an absence of contradictions in our constructions. Second, again turning to (\ref{T-q}) we try to distinguish energy distribution (\ref{I_0_dinamicM}) into pure gravitational and matter parts. Using (\ref{T_light}) in (\ref{tm_T}) in coordinates (\ref{metric_t}), taking into account (\ref{Theta_1}), (\ref{Theta_2}) and (\ref{Theta_2+}), we easily find that $t^m_{00} =\mu + P_r\l(\bar f-\Delta f\r)$. Then, for $t_{00} =t^g_{00}+t^m_{00}$ in (\ref{I_0}) we have
\bea
t^g_{00} &=& \l\{M_0 + M[v(t,r)] + \frac{C[v(t,r)]\Theta(r)}{2}\r\} \bar f(r)\delta(\bm r) - P_r[v(t,r),r]\l(2\bar f-\Delta f\r) \!,
\m{t_00_gC}\\
t^m_{00} &=&  \mu[v(t,r),r] +P_r[v(t,r),r]\l(\bar f-\Delta f\r)\,.
\m{t_00_mC}
\eea
One can see that our goal is achieved again, that is the energy density of a pure gravitational field is described by the pointlike distribution with the use of the $\delta$-function. But, unlike previous case, the energy distribution of the null liquid outside $r=0$ is accompanied (this time) by the energy density of gravitational field! Notice that such a situation is initiated by the presence of pressure of the null liquid. At last, the results (\ref{dynamicMC})-(\ref{t_00_mC}) hold in all the cases of the background: flat and (A)dS.

\subsection{The pointlike energy distribution. The $v$-time presentation}
 \m{point_v}

 Let us consider the case of the $v$-time presentation. We use the covariant properties of the field-theoretical formalism. To obtain the energy distribution in the $v$-time presentation on sections $v= \const$ on has to obtain the ${\cal I}^{0}(\bar\xi)$-component in $v$ coordinates.  We can use the coordinate transformation (\ref{v_t}) applied to the current ${\cal I}^{\alf}(\bar\xi)$ in $t$ coordinates. Then the expression (\ref{I_0}) has to be added by ${\cal I}^{1}(\bar\xi)$ in $t$ coordinates. We take $I^{10}$ that corresponds to (\ref{I_01}) with opposite sign, differentiate it with respect to $t$ in correspondence with (\ref{CL_super+}) and obtain
\be
{\cal I}^{1}(\bar\xi) =\frac{\sqrt{-\bar g}}{\k r^2}\l[2\dot M[v(t,r)] +{\dot C[v(t,r)]\Theta(r)}\r] = - {\sqrt{-\bar g}}\mu[v(t,r),r]\,.
\m{I_1_t}
\ee
Thus after coordinate transformation (\ref{v_t}) for (\ref{I_0}) and (\ref{I_1_t}) one has
\bea
{\cal I}^{0}(\bar\xi) &=&  \sqrt{-\bar g}\l[M_0 +M(v) +\half{C(v)\Theta(r)}\r] \delta(\bm r) - {\sqrt{-\bar g}}P_r(v,r)\,
\m{I_0_v}\\
{\cal I}^{1}(\bar\xi) &=&  - {\sqrt{-\bar g}}\mu(v,r)\,,
\m{I_1_v}
\eea
in $v$ coordinates. What is burst upon the eye is the absence of the term with $\mu$ in (\ref{I_0_v}). However, it was not disappeared, it is preserved in (\ref{I_1_v}). The expressions (\ref{I_0_v}) and (\ref{I_1_v}) are obtained by direct calculations in the $v$-time presentation of sections \ref{Vaidya_type} and \ref{mass_flux} as well. Thus, omitting details, we are convinced that in  the $v$-time presentation the pointlike distribution of energy with the $\delta$-function on lightlike surfaces takes a place.

\subsection{Relations of light cones}
 \m{light_cones}

Finally, we note that the requirements (i) and (ii) in Introduction have been satisfied.
%
 %
%
Concerning the requirement (iii), one has to check it directly. The light cone for the background spacetime in coordinates (\ref{metric_t}) are defined by
\be
\l.\frac{dt}{dr}\r|_{1,2} = \pm \frac{1}{\bar f}\,.
\m{cone_back}
\ee
Now, let us rewrite the metric of the general form (\ref{metric}) in the coordinates (\ref{metric_t}):
\be
d s^2 =-\l({\bar f(r)} +\Delta f\r)dt^2 - 2\frac{\Delta f}{\bar f}drdt+ \l(\frac{1}{{\bar f(r)}}- \Delta f \r) {dr^2}+ r^2d\Omega^2
\,,
\m{metric_t+}
\ee
where $\Delta f$ is defined in the general form (\ref{dynamic_pert_t}) and that is negative by our preliminary assumptions.
In this case the light cone is defined by
\be
\l.\frac{dt}{dr}\r|_{1} =  \frac{1}{\bar f}\,\frac{\bar f -\Delta f}{\bar f +\Delta f}, \qquad \l.\frac{dt}{dr}\r|_{2} = - \frac{1}{\bar f}\,.
\m{cone}
\ee
One easily finds that the physical cone (\ref{cone}) is placed inside the background cone (\ref{cone_back}) in all the cases of asymptotics. Thus, the requirement (iii) is satisfied as well.

\section{Concluding remarks and discussion}
 \m{Concluding}
\setcounter{equation}{0}

In the present paper, we have represented static and  dynamic of the Vaidya type Schwarzschild-like black hole solutions in the form of a massive point particle. As a mathematical tool the field-theoretical reformulation of GR has been applied. For all the cases of asymptotic behaviour under consideration the energy density of a pure gravitational field includes Dirac's $\delta$-function presenting the true singularity. The presence of the null liquid is reflected in constructing this singularity as well, although there is no a singularity defined by the null liquid only. The novelty is that we generalize the result in \cite{Petrov_2005,Petrov_2018} to black holes with (A)dS asymptotics. The results are enriched by including the null liquid that is also the novelty.

In the case of absence of the null liquid the presented description is quite analogical to the situation with description of the point particle in the Newtonian gravity. Indeed, in the last case the gravitational potential is defined as $\Phi = M/r$, whereas in the present consideration the potential defined by (\ref{varkappa_bar}) and (\ref{static_tot}) is $\varkappa_{00} = 2M_0/r$. Then, for the Newtonian case the mass distribution is $\varepsilon = M\delta ({\bm r})$, whereas in the present consideration it is (\ref{I_0_static}): $\varepsilon = \sqrt{\bar g}M_0\delta ({\bm r})$.

Let us discuss the presence of the null liquid. With absence of the pressure $P_r$ the energy density of the null liquid outside the singularity is defined in the form (\ref{t_00_m+}). Then, the energy density of the gravitational field is concentrated at the singularity only, see (\ref{t_00_g+}), although the influence of the null liquid presents. With presence of the pressure $P_r$ the energy density of the null liquid outside the singularity is given in (\ref{t_00_mC}). This time, the energy density of the gravitational field concentrating at the singularity is defined outside it as well, see (\ref{t_00_gC}). In all the cases there are no singularities in ${\bm t}^m_{00}$.

The main characteristics of the Schwarzschild-like black holes under consideration are mass and mass flux only, and these characteristics could be measured only. To calculate them a minimum of data can be used. It is necessary the energy density that presented by the component ${\bm t}_{00}$ only, and the energy flux density that can be calculated by the time differentiating ${\bm t}_{00}$.
Therefore we do not consider other components ${\bm t}_{\mu\nu}$.

Returning to our earlier works \cite{Petrov_Narlikar_1996,Petrov_2005}, note that we have used various field configurations to describe the true singularity by a $\delta$-function. Some of them and related energy distributions have breaks, for example, at horizons. The advantage of the results here (and in \cite{Petrov_2018,Petrov_Pitts_2019}) is that analogous problems do not appear at all.

At last, we would remark the following. One of the goals of the present paper is to describe the true singularity (that takes a place for aforementioned types of black holes) with the use of the Dirac $\delta$-function. Such a formalism gives a hope to use well defined technique applied to generalized functions with clear mathematical properties. It is an advantage with respect to the standard description of true singularities where both mathematical and physical uncertainties are problems. However, last time a very intensive development in construction and study of so-called regular black holes, where such uncertainties are avoided, is observed. In this relation one can see, for example, \cite{Visser_Ko_2018,Zerbini_2022}, the recent book \cite{Bambi_2023} and numerous references therein. From this point of view it would be interesting to consider a possibility to apply the field-theoretical formalism to study models of regular black holes as well. We discuss it briefly.

Let us consider static spherically symmetric regular solutions. There are  many variants of such solutions. As an example, we turn to the metric:
\be
ds^2= -(1-{2m(r)}/{r})dt^2 + \frac{dr^2}{1-{2m(r)}/{r}} + r^2d\Omega^2
\m{RBH_1}
\ee
that, in spite of simplicity, unites various approaches.
To avoid a central (true) singularity one has to assume at $r\goto 0$ the behaviour $m(r)= C r^3+O(r^4)$, $m'(r)= 3C r^2+O(r^3)$, $m''(r)= 6C r+O(r^2)$ with $C$ being a constant. By this, important curvature invariants become regular at $r\goto 0$. In the frame of general relativity such a behaviour is reached in various ways, for example, by introducing a special liquid with regularized $\delta$-function in its energy density at $r\goto 0$, or, with special fields, see for a detail, for example, \cite{Visser_Ko_2018,Zerbini_2022}.

To apply the field-theoretical formalism one has to choice a background. Because (\ref{RBH_1}), as a rule, is asymptotically flat it is reasonable to choose a flat background with metric (\ref{RBH_1}) when $m(r)=0$. Then, one needs to construct a current (\ref{current_Lambda}) where at the right hand side the total energy-momentum ${\bm t}_{\mu\nu}$ is defined both by metric perturbations and by a special liquid (or fields). However it is not necessary to use the structure of ${\bm t}_{\mu\nu}$ to define the current (although, in a concrete case a detail study of the structure of ${\bm t}_{\mu\nu}$ could be important). In practice, to find components of the current one calculates the superpotential components by (\ref{C_Law+}) and uses the divergence in (\ref{CL_super+}). Then, due to the remarked behaviour of $m(r)$, $m'(r)$ and $m''(r)$ one can see that for the solution (\ref{RBH_1}) the current (\ref{current_Lambda}) has no any singularity at $r=0$.

However, the use of the initial metric in the Schwarzschild-like form (\ref{RBH_1}) leads to breaks in energy distribution at horizons analogously to our earlier works \cite{Petrov_Narlikar_1996,Petrov_2005}. To avoid this problem one has to use the initial metric in the Eddington-Finkelstein-like form (\ref{metric}):
\be
ds^2= -(1-{2m(r)}/{r})dv^2 +2dvdr + r^2d\Omega^2
\m{RBH_2}
\ee
and follow the prescription of the present paper. Usually, regular black hole solutions have two or more horizons \cite{Visser_Ko_2018}. Because $m(r)$ is assumed to be enough smooth we can estimate that, at least, in the case of two horizons the energy densities related to a construction on (\ref{RBH_2}) have no breaks. Thus, as it is anticipated, in the field-theoretical description the model corresponding to (\ref{RBH_2}) is not a point-like one, it is to be rather spherically symmetric regular prolonged ``body-like'' object.

Summing up, it is not necessary to resolve a singularity problem by application of the field-theoretical formalism  to regular black holes because these problems are resolved from the start under the construction of such models. However, because diversity and multiplicity of regular black holes are huge, application of the field-theoretical formalism could be interesting in any cases to clarify or accent important properties of concrete solutions. For example, there are models with a collapse to regular black holes, so in \cite{Malafarina_2023} a collapse of dust is considered. Already, in the framework of the field-theoretical approach the collapse of dust to the usual Schwarzschild black hole has been studied \cite{Petrov_2018}. It could be interesting to represent the solution \cite{Malafarina_2023} in the field-theoretical form and compare the results with those in \cite{Petrov_2018}. We plan to do it in future.


\subsection*{Acknowledgments} The author is very grateful to Brian Pitts and George Alekseev for reading the manuscript and useful comments, he also very thanks Aleksei Staroboinsky, Igor Bulygin and Olga Sazhina for the explanation of their articles.
The work has been supported by the Interdisciplinary Scientific and Educational School of Moscow University “Fundamental and Applied Space Research”.

\subsection*{Data Availability Statement} This manuscript has no associated data
or the data will not be deposited. [Authors’ comment: All calculations
have been presented directly in the text.]

\bibliography{references}
\bibliographystyle{Style}

\ed